\documentclass[pra,byrevtex,superscriptaddress,twocolumn]{revtex4}%
\usepackage{amsfonts}
\usepackage{amsmath}
\usepackage{amssymb}
\usepackage{graphicx}%
\begin{document}
\title{All-optical reconstruction of atomic ground-state population}
\author{P. London}
\affiliation{Department of Physics, Technion-Israel Institute of Technology, Haifa 32000, Israel}
\author{O. Firstenberg}
\affiliation{Department of Physics, Technion-Israel Institute of Technology, Haifa 32000, Israel}
\author{M. Shuker}
\affiliation{Department of Physics, Technion-Israel Institute of Technology, Haifa 32000, Israel}
\author{A. Ron}
\affiliation{Department of Physics, Technion-Israel Institute of Technology, Haifa 32000, Israel}


\begin{abstract}
The population distribution within the ground-state of an atomic
ensemble is of large significance in a variety of quantum optics
processes. We present a method to reconstruct the detailed
population distribution from a set of absorption measurements with
various frequencies and polarizations, by utilizing the
differences between the dipole matrix elements of the probed
transitions. The technique is experimentally implemented on a
thermal rubidium vapor, demonstrating a population-based analysis
in two optical pumping examples. The results are used to verify
and calibrate an elaborated numerical model, and the limitations
of the reconstruction scheme which result from the symmetry
properties of the dipole matrix elements are discussed.

\end{abstract}
\maketitle

\section{Introduction}

In the past decades, alkali-metal atoms have become a central ingredient in
quantum optics applications, mostly due to the simplicity of their electronic
internal structure. In many atom-photon phenomena and particularly in
multi-photon processes, a key element is the distribution of populations among
the various states within the alkali ground-level manifold. As an example,
consider the implementation of a frequency reference based on coherent
population trapping (CPT), in which two resonant fields are used to create and
interrogate the coherence between two specific ground-level states
\cite{Arimondo96,knappe1460}. Dilution of the CPT states due to optical
pumping and the redistribution of population towards highly polarized states
pose a major limit on the CPT contrast \cite{VanierPRA2003}. In addition,
schemes for narrow CPT resonances have been proposed in which spin-exchange
decay is suppressed by the use of a proper population distribution
\cite{PhysRevLettHapper2004}. For such realizations of CPT and similar
physical arrangements, the detailed electronic population distribution is of
great importance.

Population reconstruction is a measurement of the projection of an unknown
density matrix along a certain quantization axis (diagonal elements). If
similar measurements are carried out from sufficient directions, a tomographic
reconstruction of the full density matrix (off-diagonal elements) is possible
\cite{PhysRevLettJessen2001}. Various reconstruction schemes have been
demonstrated in recent years
\cite{PhysRevLettSmithey93,PhysRevLettMukamel95,PhysRevLettWineland96,PhysRevLettKubinec98,PhysRevLettJessen2001,PhysRevAPeng2007,SilberfarbPRL2005,smithPRL2006}%
, offering diverse tools for the characterization of quantum
optics processes and for the verification or calibration of
theoretical models for atom-photon interactions. In many cases,
the measurements themselves or the rotations of the measured axes
perturb or even erase the measured state. In other cases, the
requirements of the probing protocol contrast with the operating
conditions of the measured system, \textit{e.g.} an application of
a magnetic field in a magnetically-sensitive arrangement
\cite{RomalisSERF2002}. A continuous monitoring of the state
dynamics requires an interrogation method that does not
substantially damage it, which is possible for non-radiating
systems using weak all-optical probing. Recently, all-optical
measurements of dark-resonance spectra have been used to extract
the ground-state populations of cold rubidium atoms
\cite{PhysRevAPeng2007}. However, since dark resonances are
two-photon transitions requiring non-linear pumping of a certain
dark state, they cannot be carried out with very weak probe beams.
Here we propose an all-optical linear interrogation method with
essentially no lower bound for the probe intensity. Having a
negligible effect on the ensemble, the procedure can be carried
out side by side with other processes.

In \cite{ShukerPRA2008},\ re-population processes within the ground-state
manifold of a thermal rubidium vapor were investigated using a series of weak
light pulses of single frequency and polarization. The time-dependent
absorption was analyzed to estimate the ground-state decay and decoherence
rates. The scheme presented here generalizes this method for multiple
frequencies and polarizations, in order to characterize the detailed sub-level
population distribution within the ground manifold.

The populations of the different angular-mumentum states within the ground
level of an atomic ensemble can be readily measured in a spectroscopic
measurement, only when the Zeeman and the hyperfine splittings are both larger
than the spectral widths of the transitions. In a room-temperature atomic
vapor, the optical Doppler broadening is on the order of hundreds of MHz, and
the collisional (pressure) broadening in a typical case of a buffer-gas
environment may even exceed that. Thus, the magnetic lines are resolved only
for magnetic fields much larger than 100 Gauss, which are not suitable in many
setups, such as storage-of-light \cite{LukinRMP2003}, atomic frequency
references \cite{Cyr1993,knappe1460}, magnetometry \cite{schwindt:6409}, and
optical pumping \cite{HapperRMP1972}. In these situations the angular-mumetum
states are degenerate with respect to the measurement. We show that this
effective degeneracy may be lifted utilizing the differences in the transition
strengths for each state and each polarization. Therefore in contrast to
previous schemes, the reconstruction can be carried out even when the
contribution of any single state in each measurement could not be
distinguished, and thus no application of a significant external magnetic
field is required. The basic concepts of the reconstruction method are
outlined in Sec. II, and the experimental setup of the probing system is
described in Sec. III. The results for various optical-pumping experiments are
presented in Sec. IV.

\section{Theory of population reconstruction\label{Theory_pop_recon}}

We study optical atomic transitions of the form $\left\vert F,m_{F}%
\right\rangle \rightarrow\left\vert F^{\prime},m_{F^{\prime}}\right\rangle ,$
where $F$ and $m_{F}$ are respectively the hyperfine and Zeeman numbers within
the ground-level manifold, $F^{\prime}$ and $m_{F^{\prime}}$ denote the
excited manifold. For the sake of simplicity, we consider here only
circularly-polarized light, with $m_{F^{\prime}}=m_{F}\pm1$ corresponding to
the $\sigma^{\pm}$ polarizations. The absorption of a weak probe field due to
each atomic transition depends on the inherent strength of the transition,
determined by the electric dipole moment, and on the frequency detuning from
the resonance. The absorption coefficient is thus written as the sum over the
contributions of all possible transitions,%
\begin{equation}
\alpha^{\pm}(\omega)=%
%TCIMACRO{\tsum \limits_{\left\vert F,m_{F}\right\rangle }}%
%BeginExpansion
{\textstyle\sum\limits_{\left\vert F,m_{F}\right\rangle }}
%EndExpansion
C_{\left\vert F,m_{F}\right\rangle }^{\pm}(\omega)\times P_{\left\vert
F,m_{F}\right\rangle },\label{eq_alpha}%
\end{equation}
where
\begin{equation}
C_{\left\vert F,m_{F}\right\rangle }^{\pm}(\omega)\mathbf{=}n_{0}\frac
{4\pi\omega}{c\hbar}%
%TCIMACRO{\tsum \limits_{F^{\prime}}}%
%BeginExpansion
{\textstyle\sum\limits_{F^{\prime}}}
%EndExpansion
K_{F,m_{F}}^{F^{\prime},m_{F}\pm1}(\omega)(\mu_{F,m_{F}}^{F^{\prime},m_{F}%
\pm1})^{2}\label{eq_C}%
\end{equation}
is the excitation spectrum of a specific ground-level state ($\left\vert
F,m_{F}\right\rangle $), $P_{\left\vert F,m_{F}\right\rangle }$ is its
population, and $n_{0}$ is the atomic density. The dipole matrix element is
expressed in terms of the Clebsch-Gordan coefficients as
\cite{SteckRb87LineData}
\begin{align}
\mu_{F,m_{F}}^{F^{\prime},m_{F}\pm1}  & =\mu_{0}\sqrt{2F^{\prime}+1}\left\{
\begin{array}
[c]{ccc}%
J & J^{\prime} & 1\\
F & F^{\prime} & I
\end{array}
\right\}  \times\nonumber\\
& \langle F,m_{F},1,\pm1|F^{\prime},m_{F}\pm1\rangle,
\end{align}
where $\mu_{0}$ factors out the coefficients common to all transitions, $J$
and $J^{\prime}$ are the total electronic angular momenta of the ground and
excited levels, respectively, and $I$ is the nuclear angular momentum. The
frequency dependence of the spectrum is introduced via the Voigt functions,
$K_{F,m_{F}}^{F^{\prime},m_{F}\pm1}(\omega)$, centered around the resonance
frequencies of $\left\vert F,m_{F}\right\rangle \rightarrow\left\vert
F^{\prime},m_{F^{\prime}}\right\rangle $ and accounting for both the
homogeneous and the inhomogeneous broadening components. Sampling the spectra
of the $\sigma^{+}$ and $\sigma^{-}$ polarizations at $N$ different
frequencies transforms Eq. (\ref{eq_alpha}) into a matrix form,%

\begin{equation}
\mathbf{\alpha}_{\left(  2N\times1\right)  }=\mathbf{\bar{C}}_{\left(
2N\times M\right)  }\cdot\mathbf{P}_{\left(  M\times1\right)  }\mathbf{,}%
\label{mainsysofeqn}%
\end{equation}
where each row corresponds to a single measurement, $\mathbf{\alpha}$ is the
vector of measured absorption coefficients, $\mathbf{P}$ is the ground-state
populations, and $\mathbf{\bar{C}}$ is the coupling matrix obtained from
Eq.\ (\ref{eq_C}). The population reconstruction is thus possible via a matrix
inversion, providing that at least $M$ linearily-indpendent measurements are
made. A least-squares fit to the data with $2N\gg M$ measurements would
increase the accuracy and robustness of the reconstruction, and is more
adequate in the presence of real-life measurement noise and systematic errors.
Such a fit to Eq.(\ref{mainsysofeqn}) is given by
\begin{equation}
\mathbf{P}=\mathbf{\bar{C}}^{+}\mathbf{\alpha},\label{regressedeqn}%
\end{equation}
where $\mathbf{\bar{C}}^{+}=(\mathbf{\bar{C}}^{\mathbf{T}}\mathbf{\bar{C}%
})^{-1}\mathbf{\bar{C}}^{\mathbf{T}}$ is the Moore-Penrose pseudoinverse
\cite{NumericalRecipes1992}. Finally, if the atomic density, $n_{0}$, is not
known \textit{a priori} within the desired accuracy, it should be determined
by the constrain $%
%TCIMACRO{\tsum }%
%BeginExpansion
{\textstyle\sum}
%EndExpansion
P_{\left\vert F,m_{F}\right\rangle }=1$. In practice, this can be achieved
directly from the measured spectra using the relation%
\begin{equation}%
%TCIMACRO{\tint }%
%BeginExpansion
{\textstyle\int}
%EndExpansion
d\omega\lbrack\alpha^{+}(\omega)+\alpha^{-}(\omega)]=Qn_{0}%
\frac{4\pi\omega}{c\hbar},
\end{equation}
where the identity $%
%TCIMACRO{\tsum _{F^{\prime}}}%
%BeginExpansion
{\textstyle\sum_{F^{\prime}}}
%EndExpansion
[(\mu^{+})^{2}+(\mu^{-})^{2}]=Q$ ($Q=2/3$ for rubidium-87 D1 transitions) and $%
%TCIMACRO{\tint }%
%BeginExpansion
{\textstyle\int}
%EndExpansion
d\omega K(\omega)=1$ are used.

In order to demonstrate the strengths and the limitations of our scheme, we
hereafter concentrate on the D1 transition of rubidium-87. The ground and
upper manifolds of this transition consist of two hyperfine levels, $F=1,2$,
which in turn are Zeeman splitted, as depicted in Fig.\ref{systemsetup}b. When
the Zeeman and the hyperfine splittings are larger than the spectral
line-width, it is easy to show that $\mathbf{\bar{C}}^{\mathbf{T}}%
\mathbf{\bar{C}}$ is nearly diagonal and trivially invertible, and the
populations of the eight ground-level states are immediately obtained from
Eq.(\ref{regressedeqn}). In the opposite limit where \emph{both} splittings
are small, and the spectral lines are indistinguishable, two of the
eigenvalues of $\mathbf{\bar{C}}^{\mathbf{T}}\mathbf{\bar{C}}$ in
Eq.(\ref{regressedeqn}) are close to zero due to the Clebsch-Gordan properties
\footnote{The two eigenvalues remain small also if absorption measurements of
a $\pi$-polarized probe are added to the reconstruction procedure.}. For this
limiting case, our reconstruction scheme becomes sensitive to noise and errors
and thus not applicable.

Nevertheless in the most common \emph{intermediate} regime, the Zeeman
sublevels are degenerate but the ground and excited hyperfine sublevles are
spectrally resolved. For a thermal rubidium vapor, the ground-level hyperfine
splitting is much larger than the spectral line-width (6.8 GHz compared with
$\sim$500 MHz), and the absorption spectra of the $F=1$ and the $F=2$
manifolds are completely separable. The spectrum of each $F$ manifold, for a
given probe polarization $q$, provides two meaningful data points, $\xi
_{F}^{F^{\prime}=1,q}$ and $\xi_{F}^{F^{\prime}=2,q}$, which are the sum of
all (Zeeman) transitions between $F$ and $F^{\prime}$. A fitting process
$\alpha^{q}\left(  \omega\right)  =%
%TCIMACRO{\tsum _{F,F^{\prime}=1,2}}%
%BeginExpansion
{\textstyle\sum_{F,F^{\prime}=1,2}}
%EndExpansion
\xi_{F}^{F^{\prime},q}K_{F}^{F^{\prime}}(\omega)$, where $K_{F}^{F^{\prime}%
}(\omega)$ are the degenerate Voigt profiles, gives $\xi_{F}^{F^{\prime},q}$
with high accuracy. With $\xi_{F}^{F^{\prime},q}$ known, one is left with the
following reduced equation,
\begin{equation}
\xi_{F}^{F^{\prime},q}=%
%TCIMACRO{\tsum _{\left\vert F,m_{F}\right\rangle }}%
%BeginExpansion
{\textstyle\sum_{\left\vert F,m_{F}\right\rangle }}
%EndExpansion
(\mu_{F,m_{F}}^{F^{\prime},m_{F}\pm q})^{2}P_{\left\vert F,m_{F}\right\rangle
},\label{reducedeqn}%
\end{equation}
for each ground-state manifold. For example, using circular polarizations
$\sigma^{\pm}$, a rank-3 matrix,%
\begin{equation}
\left(
\begin{array}
[c]{c}%
\xi_{1}^{1,+}\\
\xi_{1}^{2,+}\\
\xi_{1}^{1,-}\\
\xi_{1}^{2,-}%
\end{array}
\right)  =\left(
\begin{array}
[c]{ccc}%
1/12 & 1/12 & 0\\
1/12 & 1/4 & 1/2\\
0 & 1/12 & 1/12\\
1/2 & 1/4 & 1/12
\end{array}
\right)  \left(
\begin{array}
[c]{c}%
P_{\left\vert 1,-1\right\rangle }\\
P_{\left\vert 1,0\right\rangle }\\
P_{\left\vert 1,+1\right\rangle }%
\end{array}
\right)  ,
\end{equation}
is obtained for the $F=1$ manifold, with a solution%
\begin{equation}
\left(
\begin{array}
[c]{c}%
P_{\left\vert 1,-1\right\rangle }\\
P_{\left\vert 1,0\right\rangle }\\
P_{\left\vert 1,+1\right\rangle }%
\end{array}
\right)  =3\left(
\begin{array}
[c]{ccc}%
3 & 1 & -6\\
1 & -1 & 6\\
-1 & 1 & -2
\end{array}
\right)  \left(
\begin{array}
[c]{c}%
\xi_{1}^{1,+}\\
\xi_{1}^{2,+}\\
\xi_{1}^{1,-}%
\end{array}
\right)  .
\end{equation}
For the $F=2$ manifold, the matrix obtained is also of rank 3, whereas the
manifold contains 5 sublevels, thus preventing the inversion. The addition of
other excitations, either to different levels (e.g., the D2 transitions) or
with a $\pi$-plorization, are ineffective in resolving this case. This follows
from the symmetry properties of the dipole matrix elements, namely the
absorption of an \emph{isotropic} light field \cite{SteckRb87LineData} and the
spherical behavior of the dipole moment,%
\begin{equation}
\mu_{F,m_{F}}^{F^{\prime},m_{F}-q}=\mu_{F,(-m_{F})}^{F^{\prime},(-m_{F})+q},
\end{equation}
enabling no more than three independent relations. In conclusion, the detailed
populations of the $F=1$ manifold, \textit{i.e.} $P_{\left\vert
1,-1\right\rangle },$ $P_{\left\vert 1,0\right\rangle },$ and $P_{\left\vert
1,+1\right\rangle },$ can be recovered, but the $F=2$ populations can not. The
\emph{total} population in the $F=2$ manifold is determined, and in specific
cases, as we show in what follows, the total $F=2$ population can be related
to that of the maximally-polarized states $\left\vert 2,\pm2\right\rangle $,
in effect reconstructing the complete ground-level population distribution.

\section{Experimental Setup\label{Sec_Exp_setup}}%

%TCIMACRO{\FRAME{ftbpFU}{3.2465in}{3.333in}{0pt}{\Qcb{(color online). The setup
%of experiment I: a. VCSEL---vertical-cavity surface-emitting laser diode;
%PBS---polarizing beam splitter; PLZ---GLAN polarizer; MS---magnetic shield,
%EC---electric coil; ECDL---external-cavity diode laser. b. Energy levels in
%experiment I, and exciting pump fields, showing resonant transitions in red
%solid lines and non-resonant transition in blue dashed lines. The end-state
%$\left\vert F=2,m_{F}=+2\right\rangle $ is circled. c. Energy levels in
%experiment II and exciting pump fields.}}{\Qlb{systemsetup}}{setup_ver16.eps}%
%width 3.2465in;  height 3.333in;  depth 0pt;  original-width 18.667in;
%original-height 19.1703in;  cropleft "0";  croptop "1";  cropright "1";
%cropbottom "0";  filename '../setup_ver16.eps';file-properties "XNPEU";}}}%
%BeginExpansion
\begin{figure}
[ptb]
\begin{center}
\includegraphics[
height=3.333in,
width=3.2465in
]%
{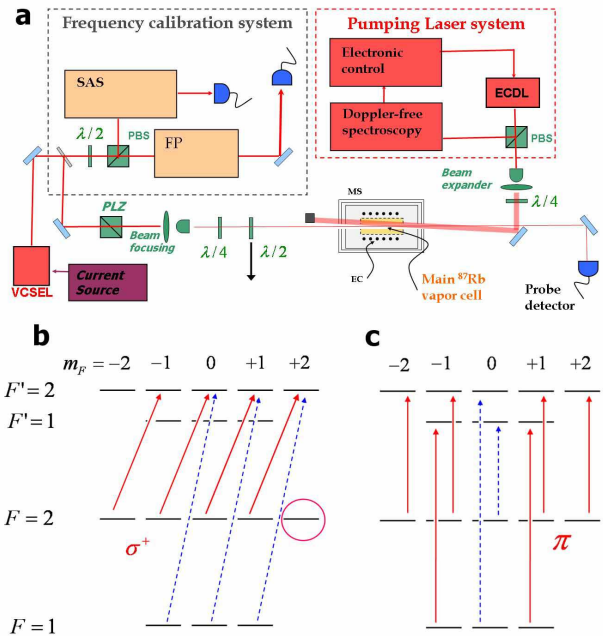}%
\caption{(color online). The setup of experiment I: a. VCSEL---vertical-cavity
surface-emitting laser diode; PBS---polarizing beam splitter; PLZ---GLAN
polarizer; MS---magnetic shield, EC---electric coil; ECDL---external-cavity
diode laser. b. Energy levels in experiment I, and exciting pump fields,
showing resonant transitions in red solid lines and non-resonant transition in
blue dashed lines. The end-state $\left\vert F=2,m_{F}=+2\right\rangle $ is
circled. c. Energy levels in experiment II and exciting pump fields.}%
\label{systemsetup}%
\end{center}
\end{figure}
%EndExpansion

The experimental arrangement, with both the pumping and the
probing setups, is depicted in Fig. \ref{systemsetup}a. We drive
the atomic ensemble to a specific atomic state using a pump beam,
as detailed in sections \ref{exp_res_exp1} and \ref{exp_res_exp2}.
Simultaneously with the pumping process, we employ a continuous
weak light field for probing the medium (Fig. \ref{systemsetup}a).
For the probe beam, we use a vertical-cavity surface-emitting
laser diode (VCSEL), and its frequency is scanned by ramping the
current. The probe beam, typically of $\sim$1mm diameter and
$\sim$50nW power, traverses a Glan-Laser polarizer followed by a
quarter-wave plate, to produce a $\sigma^{+}$ polarization. A
half-wave plate, mounted on a flip-flop, is used to convert the
probe polarization to $\sigma^{-}$ in alternating measurements.
The probe then passes through the main vapor cell and is measured
by a photo-diode. The frequency scans are of $\sim$12 GHz and
carried out at a repetition rate which is sufficiently small in
order to probe the stady-state distribution (12.5 Hz). Each scan
consists of 2500 data points (See $N$ in Eq.
(\ref{mainsysofeqn})). The population reconstruction scheme
studied here requires high accuracy determination of the frequency
of the probe. To avoid errors caused by frequency drifts of the
VCSEL, and nonlineartities in the frequency scan, a frequency
calibration system was utilized. An accurate frequency scale was
acquired by a saturated absorption spectroscopy setup
\cite{preston1432} in a reference vapor cell. Nonlinearities in
the frequency scan were corrected using frequency markers produced
by a Fabry-Perot cavity \cite{BudkerAJP1998}. After the frequency
calibration process, the absorption spectra were introduced into
Eq. (\ref{regressedeqn}) and regressed with a numerical code.

\section{Experimental results}

\subsection{Experiment I: Optical pumping with $\sigma^{+}$ polarized
light\label{exp_res_exp1}}

In the first experiment, we pump the atomic ensemble with an external-cavity
diode laser (ECDL), stabilized to the $F=2\rightarrow F^{\prime}=2$ transition
of the D1 manifold (Fig. \ref{systemsetup}b). As illustrated in Fig.
\ref{systemsetup}a, the pumping laser passes through a polarizing beam
splitter (PBS) followed by a quarter-wave plate to produce the circular
polarization. The beam is then directed into the 5-cm long main vapor cell,
containing isotopically pure $^{87}$Rb and 10 Torrs of neon buffer gas. The
cell is at room temperature, providing a vapor density of $\sim$10$^{10}$/cc.
The cell is located within a three-layered magnetic shield, and an electric
coil produces a small axial magnetic field ($B_{z}=450$mG) to set the
quantization axis.%

%TCIMACRO{\FRAME{ftbpFU}{3.6374in}{3.3762in}{0pt}{\Qcb{(color online) Measured
%signals in experiment I: a. FP signal in blue thin lines and SAS signal in red
%thick line are used for callibrating the frequency axis. (b,c,d) Absorption
%spectra measured with $\sigma^{+}$ polarization (solid lines) and $\sigma^{-}$
%polarization (dashed lines) with various pump powers, showing the thermal
%equilbrium (b), depopulated $F=2$ (c), and vacancy of the $m_{F}\leq1$ states
%in $F=2$ (d). On the right, are the population distribution of the $F=1$
%manifold and the total population of $F=2$ manifold.}}{\Qlb{abslinesimap}%
%}{abslinessigmap_subplot.eps}{\special{ language "Scientific Word";
%type "GRAPHIC";  maintain-aspect-ratio TRUE;  display "USEDEF";
%valid_file "F";  width 3.6374in;  height 3.3762in;  depth 0pt;
%original-width 5.8219in;  original-height 4.3708in;  cropleft "0";
%filename '../abslinessigmap_subplot.eps';file-properties "XNPEU";}}}%
%BeginExpansion
\begin{figure}
[ptb]
\begin{center}
\includegraphics[
height=3.3762in,
width=3.6374in
]%
{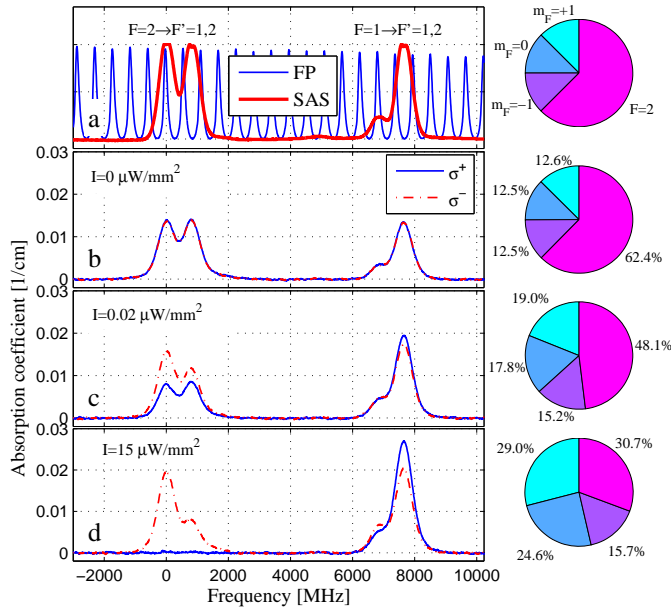}%
\caption{(color online) Measured signals in experiment I: a. FP
signal (blue thin lines) and SAS signal (red thick line) are used
for callibrating the frequency axis. (b,c,d) Absorption spectra
measured with $\sigma^{+}$ polarization (solid lines) and
$\sigma^{-}$ polarization (dashed lines) with various pump powers,
showing the thermal equilbrium (b), depopulated $F=2$ (c), and
vacancy of the $m_{F}\leq1$ states in $F=2$ (d). On the right are
the population distribution of the $F=1$ manifold and the total
population of
$F=2$ manifold.}%
\label{abslinesimap}%
\end{center}
\end{figure}
%EndExpansion
We have performed population-distribution measurements with this arrangement,
as described in Sec. \ref{Sec_Exp_setup}, for various pump powers. Figure
\ref{abslinesimap} depicts probe absorption curves for three typical pump
powers. When no pump is applied (panel b in Fig. \ref{abslinesimap}) the
medium is expected to be in thermal equilibrium, in which the population is
equally distributed among the eight levels of the ground-state. Indeed, the
measured $F=1$ populations are equal up to the measurment error. As the pump
power is increased, the absorption from the $F=2$ manifold is reduced due to
the depopulation of the levels within the $F=2$ manifold. The $\sigma^{+}$ and
$\sigma^{-}$ absorption curves differ as a consequence of the increasing
polarization of the medium. For pump intensities larger than 10$\mu$W/mm$^{2}%
$, the $\sigma^{+}$ absorption from the $F=2$ manifold nullifies, showing that
the $\left\vert F=2,m_{F}=-2,-1,0,+1\right\rangle $ levels are vacant.

The populations of the three states within the ground-state $F=1$
manifold are reconstructed using the procedure described in Sec.
\ref{Theory_pop_recon}. The best-fitted values of the homogeneous
and inhomogeneous broadenings of the Voigt lineshape,
$\Gamma=$103~MHz and $\sigma=$202 MHz (FWHM $\sim$600 MHz), are as
expected for our room-temperature buffer-gas system. Although the
population distribution of the $F=2$ manifold is not formally
solved, a least-squares minimization process assuming non-negative
values can be used to estimate the detailed populations within the
$F=2$ manifold. At high pump powers, in which the population of
$F=2$ is associated with $\left\vert F=2,m_{F}=+2\right\rangle $
only, this estimation becomes reliable and the populations of
\emph{all} eight states within the ground-level are extracted. For
the strongest pump used in our experiment, over $95\%$ of the
$F=2$ population piles up in the $m_{F}=+2$ level. The
reconstructed population distribution is shown in the inset of
each panel in Fig. \ref{abslinesimap}.

\begin{figure}
[ptb]
\begin{center}
\includegraphics[
height=2.7406in,
width=3.6417in
]%
{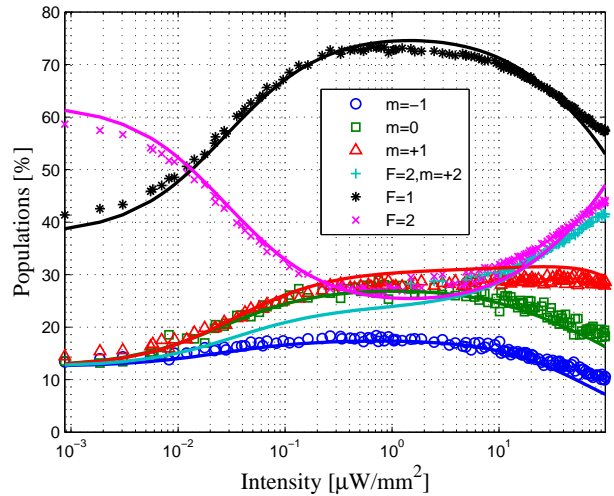}%
\caption{The reconstructed populations versus five decads of the pump
intensity in experiment I (symbols). The numerical model (lines) agrees well
with the measurements.}%
\label{spintempresults}%
\end{center}
\end{figure}

The detailed population measurements provide a substantial dataset, which can
be used to benchmark a theoretical multilevel model \footnote{Our numerical
model describes the dynamics and the steady-state solutions of the atomic
density-matrix, within a Master-equation formalism, and was previously used in
\cite{ShukerPRA2008,RanOptExp2009}. The model incorporates all relevant
sublevels, \textit{e.g.} the 16 sublevels of the rubidium-87 D1 transition,
and resonant coupling fields. Decay and decoherence processes, within the
ground and the excited manifolds and between them, are effectively described
by a suitable Lindblad form. We use 11 velocity groups to model the doppler
broadening.}. In Fig. \ref{spintempresults}, we show a comparison between the
experimental measurements and the results of our multilevel numerical model,
for a wide range of pump powers. The fitted Voigt parameters of the measured
spectra ($\Gamma$ and $\sigma$) were used as an input to the model. A single
model parameter that determines the ground-level population equilibration was
fitted to the data, effectivly describing the diffusion-induced decay and wall
collisions (decay due to spin-exchange collisions is negligible in our
conditions). It is evident that our model describes well the population
distribution within the ground levels for five decades of pump power.%
%TCIMACRO{\FRAME{ftbpFU}{3.6417in}{2.7406in}{0pt}{\Qcb{The reconstructed
%populations versus five decads of the pump intensity in experiment I
%(symbols). The numerical model (lines) agrees well with the measurements.}%
%}{\Qlb{spintempresults}}{result_ft2.eps}%
%width 3.6417in;  height 2.7406in;  depth 0pt;  original-width 5.8219in;
%original-height 4.3708in;  cropleft "0";  croptop "1";  cropright "1";
%cropbottom "0";  filename '../result_FT2.eps';file-properties "XNPEU";}}}%
%BeginExpansion
%EndExpansion

When the pump power is very weak a thermal distribution within the
ground-state is observed. As the pump power is increased, the medium is
polarized and more population accumulates in the higher $m_{F}$ levels of the
$F=1$ manifold (triangles and squares in Fig. \ref{spintempresults}). At
higher pump powers, the total population of the $F=1$ manifold is depleted due
to off-resonant pumping. All these phenomena are well described by the
multilevel model with a single fit parameter. Remaining discrepancies result
from magnetic field inhomogeneity and from residual $^{85}$Rb isotope in the
cell (about 2$\%$). It\ is interesting to note\ that an extremely efficient
pumping to the maximally polarized state, $\left\vert F=2,m_{F}=2\right\rangle
$, can be achieved by tuning the frequency of the pumping laser in between the
ground-level hyperfine transitions. We verified this experimentally by locking
the laser to the $F=3\rightarrow F^{\prime}=2 $ transition of $^{85}$Rb, and
found that more than 90\% of the ground-level population was pumped to the
maximally polarized state for a pump intensity of about 100 $\mu$W/mm$^{2}$.

\subsection{Experiment II: Optical pumping with $\pi$ polarized
light\label{exp_res_exp2}}

We now turn to implement our reconstruction method for the case of CPT-based
atomic frequency reference. One of the well known limitations of these
frequency references is optical pumping of the atomic population from the
clock-states $\left\vert F=1,m_{F}=0\right\rangle $ and $\left\vert
F=2,m_{F}=0\right\rangle $ \cite{VanierPRA2003}. An approach previously
investigated to resolve this problem is \ the repopulation of the clock states
using additional $\pi$-polarized light field
\cite{MatisovEJPD2005,AsifMaster2009}. To study this repopulation technique,
we pump the ensemble with a RF-modulated $\pi$-polarized field, as depicted in
Fig. \ref{systemsetup}c. We use a current-modulated VCSEL at a modulation
frequency of 3.0 GHz. The laser frequency is tuned such that the +1 and -1
sidebands are resonant with the $F=1\rightarrow F^{\prime}=1$ and
$F=2\rightarrow F^{\prime}=2$ transitions, respectively. The laser is linearly
polarized with a PBS and directed into a vapor cell containing isotopically
pure $^{87}$Rb and 10 Torr of nitrogen buffer gas, at temperature of 58$^{0}%
$C. The vapor cell is magnetically shielded and a constant magnetic field of
about 100mG is applied to set the quantization axis along the pump beam
polarization direction (to induce $\pi$ transitions). The probe beam, used for
the population distribution measurement, is applied along the magnetic field axis.

\begin{figure}
[ptb]
\begin{center}
\includegraphics[
height=2.9354in, width=3.9in
]%
{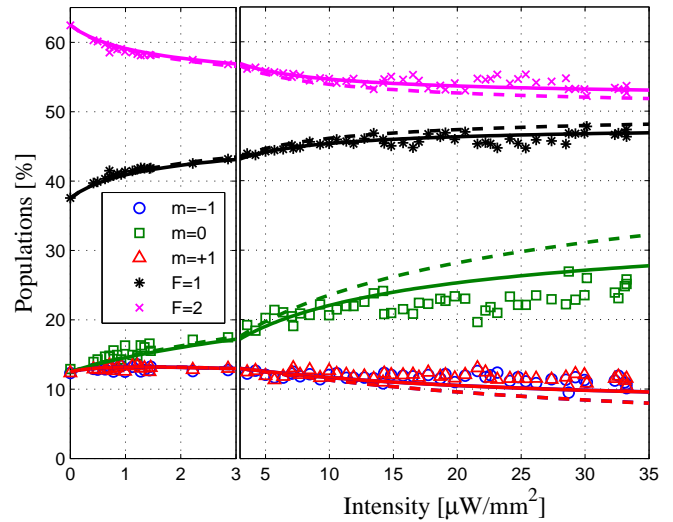}%
\caption{(color online) The reconstructed populations versus the
pump intensity in experiment II, showing the agreement between the
measured population (symbols), a numerical model (solid lines) and
a numerical model that neglects
non-resonant transitions (dashed lines) }%
\label{piresults}%
\end{center}
\end{figure}

Fig. \ref{piresults} presents the reconstructed populations as well as the
calculated results of our multilevel model.
%TCIMACRO{\FRAME{ftbpFU}{3.6365in}{2.7371in}{0pt}{\Qcb{(color online) The
%reconstructed populations vesus the pump intensity in experiment II, showing
%the agreement between the measured population (symbols), a numerical model
%(solid lines) and a numerical model that neglects non-resonant transitions
%(dashed lines) }}{\Qlb{piresults}}{pi_pumping_3ghz_mod.eps}%
%width 3.6365in;  height 2.7371in;  depth 0pt;  original-width 5.8219in;
%original-height 4.3708in;  cropleft "0";  croptop "1";  cropright "1";
%filename '../pi_pumping_3GHz_mod.eps';file-properties "XNPEU";}}}%
%BeginExpansion
%EndExpansion
The $\left\vert F=1,m_{F}=0\right\rangle $ clock-state accumulates excess
population as the pump power is increased, reaching a maximal population of
twice the thermal equilibrium population. The populations in the $m_{F}=\pm1$
states show symmetric behavior, demonstrating that a pure $\pi$-pumping was
applied to the sample. As the pump power is increased, the population of the
$F=2$ manifold decreases, since the $F=2\rightarrow F^{\prime}$ transitions
are more efficient than the $F=1\rightarrow F^{\prime}$ transitions. To
emphasize the destructive effect of the non-resonant excitations (blue dashed
arrows in Fig. \ref{systemsetup}c), we have performed an additional
calculation, in which the non-resonant excitations were excluded (dashed lines
in Fig. \ref{piresults}). It is evident that the non-resonant excitations pose
a limit on the efficiency of the $\pi$ re-pumping technique.

\section{conclusions}

We have presented an all-optical scheme to measure the populations
within the ground-state of multilevel atoms and have demonstrated
its implementation for a $^{87}$Rb vapor. In our scheme, the
required light and magnetic fields are weak, and the degeneracy
created by the optical-transitions width is lifted by utilizing
the properties of the dipole matrix elements. As an intuitive
quantity characterizing the state of the ensemble, the population
distribution is most natural in monitoring dynamic processes, and
in analyzing the initial quantum state in precise experiments
\cite{KuzmichPRL2006}. For the D1 transition of $^{87}$Rb, the
scheme reconstructs the detailed population distribution of the
$F=1$ manifold and the total population of the $F=2$ manifold. In
the framework of linear spectroscopic measurements, the population
distribution of the $F=2$ level is fully reconstructed only if one
can relate the total population of the $F=2$ manifold to one of
the maximally polarized states. In order to reconstruct the full
density matrix, a tomographic scheme (as in
\cite{PhysRevLettJessen2001}) could be applied. For that, the
population distribution measurements, as presented in this work,
should be carried out along $4F+1$ different directions. We have
used the scheme in two optical-pumping experiments and
demonstrated its strength in describing various optical processes,
such as resonant and non-resonant excitations and depolarization,
and in verifying and calibrating the parameters of an elaborated
numerical model. We further employed the calibrated model to
calculate the improvement of the contrast of a CPT based frequency
reference.

\bigskip

\begin{center}
$\mathbf{Acknowledgments}$
\end{center}

We wish to acknowledge the helpful discussions with Amnon Fisher, Nir
Davidson, Ran Fischer and Asif Sinay. We thank Yoav Erlich for technical support.

\bibliographystyle{apsrev}
\bibliography{references}

\end{document}